\documentclass[twocolumn,epjc3]{svjour3}\pdfoutput=1
\smartqed
\usepackage{amsmath}
\usepackage{amssymb}
\usepackage{float}
\usepackage{graphicx}
\usepackage{tabularx,ragged2e,booktabs,caption}
\usepackage{caption}
\usepackage{subcaption}
\usepackage{hyperref}
\usepackage{slashed}
\usepackage{widetext}
\usepackage{adjustbox}
\RequirePackage{graphicx}
\journalname{Eur. Phys. J. C}
\usepackage{bm}

\newcommand{\bea}{\begin{eqnarray}}
\newcommand{\eea}{\end{eqnarray}}

\begin{document}
\title{Confronting Einstein Yang Mills Higgs Dark Energy in light of observations}
\author{Debabrata Adak 
\thanksref{e1,addr1}} 
\thankstext{e1}{e-mail: debabrata.adak.sinp@gmail.com}

\institute{Department of Physics,
	Government General Degree College Singur, Jalaghata Singur, Hooghly-712409,
West Bengal, India \label{addr1}}
\date{Received: date / Accepted: date}

\maketitle

\begin{abstract}
	We study the observational aspects of Einstein Yang Mills Higgs Dark energy 
	model and constrain the parameters space from the latest observational data
	from type Ia supernovae, observational Hubble data, baryon acoustic 
	oscillation data and cosmic microwave background radiation shift parameter 
	data. It is found from the analysis of data that the Higgs field in presence
	of gauge fields
	can successfully describe the present accelerated expansion of the universe
	consistent with the astrophysical observations.
	\keywords{Dark Energy \and Higgs field \and Cosmology \and Observations}
\end{abstract}

\section{Introduction}
Cosmic acceleration discovered more than two decades ago by supernova projects
\cite{Riess:1998cb,Perlmutter:1998np} is perhaps  the most important and fascinating 
phenomenon that still remains
in mystery. %
%
This cosmic acceleration can be accounted for by invoking the presence of some exotic
fluid dubbed dark energy with negative pressure to overcome the gravitational collapse 
and thereby resulting in the accelerated expansion of the universe
\cite{Caldwell:2009ix,Frieman:2008sn,RevModPhys.75.559,Linder_2008,Bamba:2012cp}.
From the observations it is evident that it constitute about 68\% of
the total energy density in the universe \cite{Aghanim:2018eyx}.
The cosmological constant $\Lambda$ is the
best fitted model so far to explain this recent accelerated expansion of the universe. 
However it suffers from two major theoretical 
problems known as 
fine tuning problem \cite{amendola2010dark} and cosmic coincidence problem 
\cite{PhysRevLett.80.1582}. 
Despite being consistent with
the observations, these problems of cosmological constant make the cosmologists
search for alternatives.

A simplest alternative is the canonical scalar field model known as ``quintessence''
\cite{PhysRevLett.80.1582}. The potential of the canonical scalar
field is so chosen that the field rolls very slowly at the present epoch resulting 
in the negative pressure of the field which leads to cosmic acceleration. This essentially
requires  the potential to be very flat with respect to the field $\phi$ resulting in the
mass of the field around $10^{-33}$ eV.
The tracker behaviour of the scalar field model \cite{PhysRevD.59.123504}
helps to alleviate the
problem of cosmic coincidence in the dark energy scenario .
Explaining the late time cosmic acceleration is also possible from the modification 
of gravity at the large scales known as infrared modification of gravity.
It is found that 
higher order curvature invariants play an important
role in modification of gravity at large scales
thereby leading to accelerated expansion of the universe at the present epoch
\cite{Capozziello:2002rd,Capozziello:2003tk,Capozziello:2003gx}.
Moreover %
higher dimensional models of gravity induces
modification of Einstein's gravity in the 3+1 dimensional effective theory 
at large scales leading to the 
accelerated expansion of the universe \cite{Nicolis:2008in,DeFelice:2010pv}.
A large number of modified gravity models is tested to be free from ghost or tachyon
instabilities and they also do not conflict with the solar system constraints
(see \cite{Clifton:2011jh,Nojiri:2017ncd} and references therein for a 
review on cosmology 
driven by modified gravity models). Of late the detections of GW170817 and GRB 170817A
revealed the fact that the speed of gravitational waves differs from the speed of light
by one part in $10^{15}$
\cite{PhysRevLett.119.161101,Goldstein_2017,Savchenko_2017,Abbott_2017a,Abbott_2017b}. 
This discovery have
severely constrained these modified gravity models as well as other dark energy models  
\cite{PhysRevLett.119.251301,PhysRevLett.119.251302,PhysRevLett.119.251303,PhysRevLett.119.251304,Creminelli_2018,Creminelli_2019,Noller:2020afd,Casalino:2018tcd,Casalino:2018wnc}.

From the standard model of particle physics this is well known that all the 
particles in the universe get masses due to their  interaction with the Higgs 
field \cite{weinberg_1996,das2008lectures}. 
The dynamics of FRW universe was studied in presence of non-abelian gauge fields 
invariant under SO(3) and SU(2) gauge group or an arbitrary gauge group SO(N)
\cite{Henneaux:1982vs,GALTSOV199117,Moniz:1990hf,Darian:1996mb,Barrow:2005df,Banijamali:2011ep,Elizalde:2012yk,Cembranos:2012ng}.
In the context of inflation Einstein Yang-Mills Higgs action was
first introduced \cite{Moniz_1993,Ochs:1996yr} to study the effect of gauge field
on inflation.
There are also models of inflation known as 
Gauge-flation where the inflation is driven 
solely by a non-Abelian gauge field, 
see for instance Refs. 
\cite{Maleknejad:2011jw,Maleknejad:2012fw,Namba:2013kia}.
In these models the Yang-Mills gauge sector
is modified to include a specific higher order 
derivative operator
for driving inflationary dynamics
whereas 
in the other works \cite{Banijamali:2011ep,Elizalde:2012yk,Cembranos:2012ng}, 
the Yang Mills gauge sector
is non-minimally coupled to gravity or associated with
some potential for the same reason.
In the case of standard model 
Higgs field no such thing is there \cite{Halzen:1984mc}.
Recently in \cite{Rinaldi:2014yta}, 
the dynamics of this non-abelian Higgs field coupled to gravity
was studied in the context of late time cosmic acceleration.
In the work \cite{Alvarez:2019ues}, considering the interaction in SU(2) representation
for Higgs field the authors have studied the dynamics of cosmology in Einstein
Yang-Mills Higgs to explain the recent accelerated expansion of the Universe. 
The main motivation behind considering the 
Yang Mills gauge field with the Higgs field 
lies in the mexican hat potential of the Higgs
field \cite{Rinaldi:2014yta}.
The mexican hat potential of the Higgs field
is steep enough to sustain the late time
cosmic acceleration without a very fine tuned
initial condition. The presence of Yang Mills 
gauge field in the standard model of particles
helps to alleviate the fine tuning of the 
initial conditions by increasing the Hubble friction
and flattening the Higgs potential as we shall see
in next section.
Moreover what is not yet known is that the viability of this model in respect of 
cosmological observations. 

In the present work we study the viability of the Einstein Yang-Mills Higgs dark
energy in the context of observational data. We constrain the model from
type Ia supernova data (SNe Ia), observational Hubble data (OHD), baryon acoustic
oscillation data (BAO) and cosmic microwave background shift parameter data (CMB)
and show that the model parameter space is consistent with the cosmological
observations thus making this a viable model for dark energy to explain the 
current accelerated expansion of the universe. This paper is organised as follows.
In Sec. \ref{EYMH} we review the Einstein Yang-Mills Higgs action coupled to gravity
and the equations of motion in FRW background to study the dynamical system.
Construction of autonomous system and dynamics of cosmology is studied in 
Sec. \ref{DoC}.
In Sec. \ref{Obs-cons} we discuss the various observational data and the formalism
for analyses of those data. In this section  we also confront this Higgs 
dark energy model 
with the observational data and present the results of our data
analysis. Eventually we conclude in Sec. \ref{conc}

\section{Einstein Yang-Mills Higgs action}
\label{EYMH}
In what follows we describe the Higgs dark energy in presence of gauge field 
in background of Einstein's gravitation.
The Einstein Yang-Mills Higgs action is given by 
\cite{Rinaldi:2014yta,Alvarez:2019ues},
\bea
S&=&\int d^4x \sqrt{-g}\left(\frac{M_{\rm Pl}^2}{2} R -
\frac{1}{4} F^{\mu\nu}_a F_{\mu\nu}^a
-(D^\mu \Phi)^\dagger(D_\mu \Phi)\right.\nonumber\\&&\left.- V(\Phi)+{\cal L}_r+{\cal L}_m\right)\,,
\label{YMHaction}
\eea
where $M_{\rm Pl}$ is the reduced Planck mass given by $M_{\rm Pl}=1/\sqrt{8\pi G}$,
$g$ is the determinant of spacetime metric, 
$\Phi$ is the complex Higgs doublet invariant under SU(2) gauge symmetry, ${\cal L}_r$
is the lagrangian for radiation and ${\cal L}_m$ is the matter lagrangian.
Here $F^a_{\mu\nu}$ is the rank-2 tensor that represents the non-Abelian gauge field
and is given by
\bea
F^a_{\mu\nu} &=& \partial_\mu A^a_\nu- \partial_\nu A^a_\mu + \beta
\epsilon ^a_{bc} A^b_\mu A^c_\nu\,\,,
\eea
where $A^a_\mu$ is the gauge field, $\beta$ is the coupling of SU(2) group and
$\epsilon^a_{bc}$ is the rank 3 Levi-Civita symbol. $D_\mu$ is the gauge covariant
derivative given by
\bea
D_\mu &=& \nabla_\mu - i\beta \frac{\sigma_a}{2} A^a_\mu\,\,,
\eea
where $\nabla_\mu$ is the spacetime covariant derivative and $\sigma_a$  are the
Pauli matrices.
The complex Higgs doublet and its potential are respectively given by,
\bea
\Phi&=&\begin{pmatrix} \phi_1 +i\chi_1 \\ \phi_2 + i\chi_2 \end{pmatrix}\,\,,
\eea
where $\phi_1$, $\phi_2$, $\chi_1$, $\chi_2$ are real scalar fields and 
\bea
V(\Phi)&=& \frac{\lambda}{4} \left(\Phi^\dagger \Phi-v^2\right)^2\,\,,
\eea
where $v$ is the vacuum expectation value (VEV) of Higgs field.

It is evident from the observations \cite{Aghanim:2018eyx}
that our universe is homogeneous and isotropic on large scales. Hence the background
spacetime of the universe is described by the Friedmann Lema\^{i}tre Robertson Walker 
(FLRW) metric and is given by in spherically symmetric coordinates,
\bea
ds^2 &=& -dt^2+ a^2(t)\left(dr^2+r^2d\Omega^2\right)\,\,,
\eea
where $t$ is the cosmological time, $a(t)$ is scale factor for expanding universe
and $d\Omega^2=d\theta^2+\sin^2 \theta d\phi^2$.
The energy momentum tensor for the action in Eq. \ref{YMHaction} is given by,
\bea
T_{\mu\nu} &=& - F^a_{\mu \eta} F_{\nu a}^\eta - (D_\mu \Phi)^\dagger (D_\nu \Phi)
	-(D_\nu \Phi)^\dagger (D_\mu \Phi)\nonumber\\&& 
+ 2 \frac{\partial}{\partial g^{\mu\nu}}({\cal L}_m+{\cal L}_r)
\nonumber\,\\
&&-g_{\mu\nu}\left[-\frac{1}{4} F^{\mu\nu}_a F_{\mu\nu}^a
-(D^\mu \Phi)^\dagger(D_\mu \Phi)- V(\Phi)+{\cal L}_r+{\cal L}_m\right]\,.
\eea
The Einstein tensor $G_{\mu \nu}$ is diagonal for FLRW background spacetime and
hence the off diagonal terms of  energy momentum tensor should vanish. This
condition makes the gauge field become $A^a_\mu=\delta^a_\mu f(t)$ 
\cite{Rinaldi:2014yta} where $f(t)$
is the only degree of freedom in the gauge sector as allowed from the FLRW
spactime of the universe, $a$ is the gauge index and $i$ is the spatial index.
As discussed in \cite{Alvarez:2019ues}, 
this condition is not sufficient to avoid the non-zero
contribution to the momentum density arising from the interaction between 
the Yang-Mills field and the Higgs field. Hence another additional condition
which is required to establish the isotropy in energy momentum tensor is to fix
the gauge so that 
\bea
\Phi(t) &=& \begin{pmatrix} \phi(t) \\ 0 \end{pmatrix}\,,
\eea
where $\phi(t)$ is a real scalar field.
With these choices of gauge field and gauge fixation for the Higgs field, we obtain
\bea
H^2&=& \frac{1}{3M_{\rm Pl}^2} \left[\frac{3}{2} \frac{\dot f(t)^2}{a(t)^2}
+\dot \phi(t)^2 + \frac{3}{2} \frac{\beta^2 f(t)^4}{a(t)^4}
	+  \frac{3}{4} \frac{\beta^2 \phi(t)^2 f(t)^2}{a(t)^2}+ \right.\nonumber\\
	&&\left.V(\phi) 
+\rho_m +\rho_r\right]\,,\label{hub2}\\
\dot H &=& -\frac{1}{2M_{\rm Pl}^2}\left[2\frac{\dot f(t)^2}{a(t)^2}+
2 \dot \phi(t)^2 + 2 \frac{\beta^2 f(t)^4}{a(t)^4}+
\frac{\beta^2 \phi(t)^2 f(t)^2}{2a(t)^2}\right.\nonumber\\
&&\left.+\rho_m +\frac{4}{3}\rho_r\right]\,,
\eea
where $H$ is Hubble parameter given by $H=\dot a(t)/a(t)$ and $\rho_m$, $\rho_r$ are
the matter and radiation density respectively. The equation of motions of the
gauge and Higgs fields are respectively given by,
\bea
\ddot f(t) + H\dot f(t) +\beta^2\left[ 2\frac{f(t)^3}{a(t)^2} 
+\frac{f(t) \phi(t)^2}{2} \right] &=& 0\,,\label{gauge-pot}\\
\ddot \phi(t) + 3 H \dot \phi(t) + \frac{3 \beta^2 f(t)^2 \phi(t)}{4 a(t)^2}
+ \frac{dV(\phi)}{d\phi}&=&0\,,\label{higgs-eff-pot}
\eea
where $V(\phi)$ is given by $V(\phi)=\frac{\lambda}{4}(\phi^2-v^2)$. This is worth
mentioning here that from Eq. (\ref{gauge-pot}) it is evident that there arises
an effective potential for the gauge field with vanishing vacuum expectation value
due to the interaction between the gauge field and the Higgs field. Moreover the
same interaction leads to the effective potential of the Higgs field also as shown
in the Eq. (\ref{higgs-eff-pot}).

\section{Dynamics of cosmology}
\label{DoC}
To analyse the cosmological dynamics the following dimensionless variables are 
introduced here.
\bea 
x_1&=& \frac{\dot f}{\sqrt{2}aM_{\rm Pl}H}\,,~~~~~
y_1= \frac{\beta f^2}{\sqrt{2}a^2 M_{\rm Pl}H}\,,\nonumber\\
z_1&=& \frac{\beta f \phi}{2aM_{\rm Pl}H}\,,~~~~~
x_2= \frac{\dot \phi}{\sqrt{3} M_{\rm Pl}H}\,,\nonumber\\
y_2&=& \sqrt{\frac{V(\phi)}{3M^2_{\rm Pl} H^2}}\,,~~~~~
r= \sqrt{\frac{\rho_r}{3M^2_{\rm Pl} H^2}}\,,\nonumber\\
m&=&  \sqrt{\frac{\rho_m}{3M^2_{\rm Pl} H^2}}\,,~~~~~
w_1= \frac{\sqrt{2} a M_{\rm Pl}}{f}\,,
\eea
The subscripts 1 and 2 refers to the dimensionless variables corresponding to
gauge field and the Higgs field. With these choices the total energy density in the
universe takes the form (from Eq. (\ref{hub2})),
\bea
x_1^2+y_1^2+z_1^2+x_2^2+y_2^2+r^2+m^2&=&1\,.
\eea
The evolution equations of the autonomous system are given by
\bea
x_1^\prime &=& x_1 (q-1) - w_1(2 y_1^2+z_1^2)\,,\nonumber\\
y_1^\prime &=& y_1(2x_1 w_1+q-1)\,,\nonumber\\
z_1^\prime &=& z_1 (x_1 w_1+q)+\frac{\sqrt{3}}{2} w_1 y_1 x_2\,,\nonumber\\
x_2^\prime &=& x_2 (q-2) - z_1 w_1 (2 \alpha y_2 + \frac{\sqrt{3}}{2} y_1)\,,\nonumber\\
y_2^\prime &=& y_2 (q+1) + \alpha w_1 z_1 x_2\,,\nonumber\\
r^\prime &=& r (q-1) \,,\nonumber\\
m^\prime &=& m \left(q- \frac{1}{2}\right)\,,\nonumber\\
w_1^\prime &=& w_1(1- w_1 x_1)\,,
\eea
where the symbol prime denotes a derivative with respect to $N=\ln a$, $a$ being
the scale factor of the universe and $q$ is the deceleration parameter defined
as $q(t)= -\frac{\ddot a(t) a(t)}{\dot a(t)^2}$. In terms of the dimensionless variables
defined above the deceleration parameter takes the form
\bea 
q&=& \frac{1}{2} (1+x_1^2+ y_1^2- z_1^2+ 3x_2^2- 3y_2^2+ r^2)\,.
\eea
Here $\alpha$ is a dimensionless constant given by 
$\alpha= \sqrt{\frac{\lambda}{2\beta^2}}$. 

\begin{figure*}
  \centering
  \includegraphics[scale=1.0]{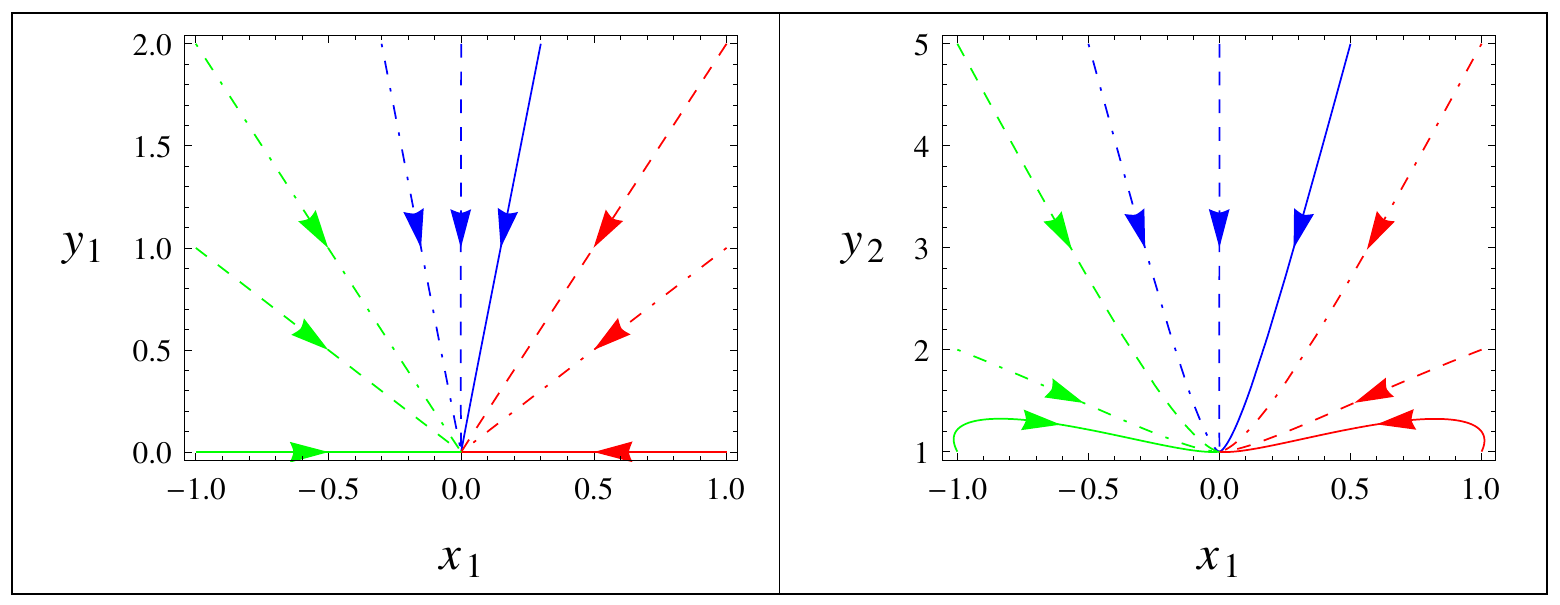}
  \caption{Evolution of the three dimensionless
  parameters $x_1$, $y_1$ and $y_2$ are shown in the 
  $(x_1,~y_1)$ and $(x_1,~y_2)$ plane. In the left figure
  the system starts from
(1, 0) (solid red), (1, 1) (red dotdashed), (1, 2) (red dashed),
 (0.3, 2) (solid blue), (0, 2) (blue dashed), (-0.3, 2) (blue dotdashed)
(-1,2) (green dotdashed), (-1,1) (green dashed) and (-1,0)
(solid green)
in the 
$(x_1,~y_1)$ plane and approaches the attractor solution
(0, 0) as $N$ changes 
from $N=-15$ to $N=0$. In the right figure the 
system starts from (1, 1) (solid red), (1, 2) (red dashed), (1, 5)
 (red dotdashed), (0.5, 5) (solid blue), (0, 5) (blue dashed),
(-0.5,5) (blue dotdashed), (-1, 5) (green dashed), (-1, 2) 
(green dotdashed) and (-1,1) (solid green) 
in the 
$(x_1,~y_2)$ plane and approaches the attractor solution
(0, 1) as $N$ changes 
from $N=-15$ to $N=0$.}
  \label{dyna}
\end{figure*}

\begin{figure}
  \centering
  \includegraphics[scale=0.6]{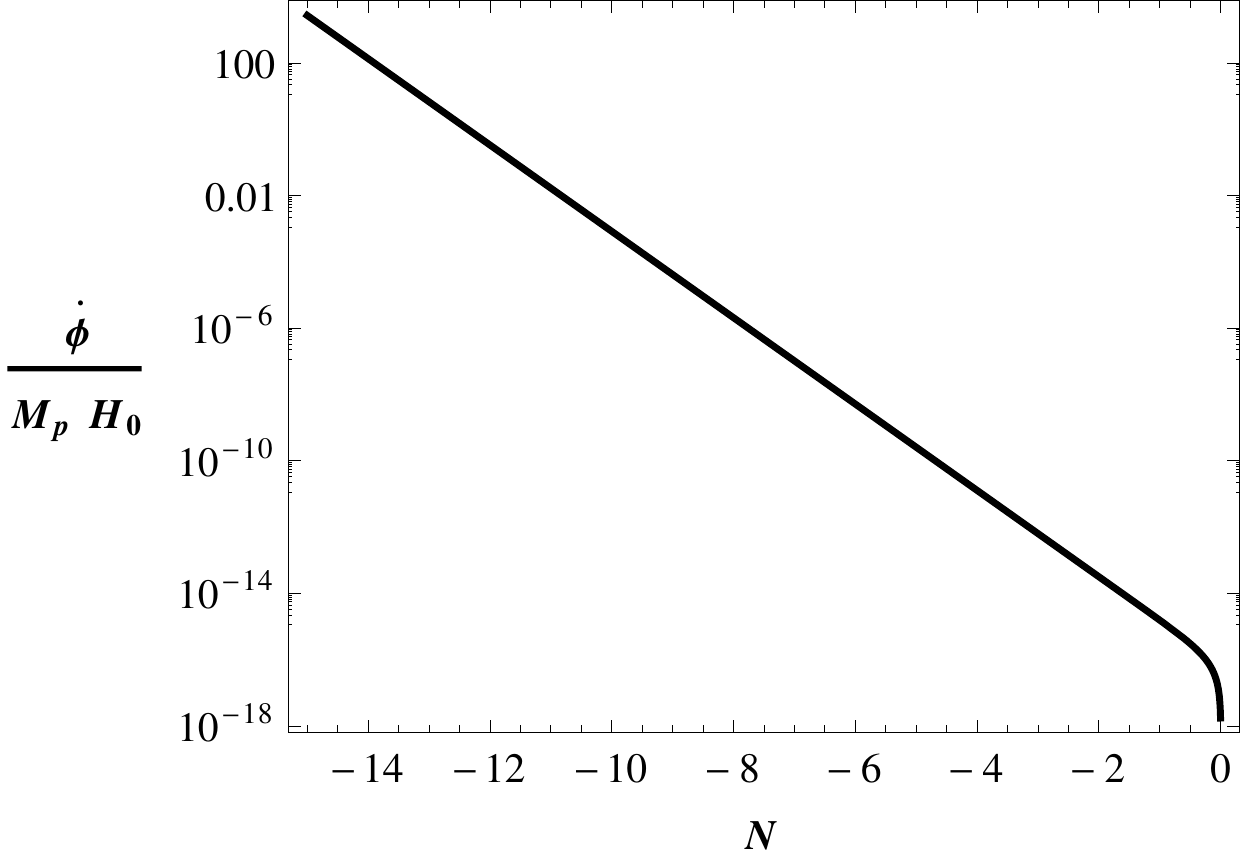}
  \caption{Velocity of Higgs field $\frac{\dot\phi}{M_P H_0}$
   plotted against the number of e-foldings $N$. }
  \label{phi-velo}
\end{figure}

\begin{figure}
  \centering
  \includegraphics[scale=0.6]{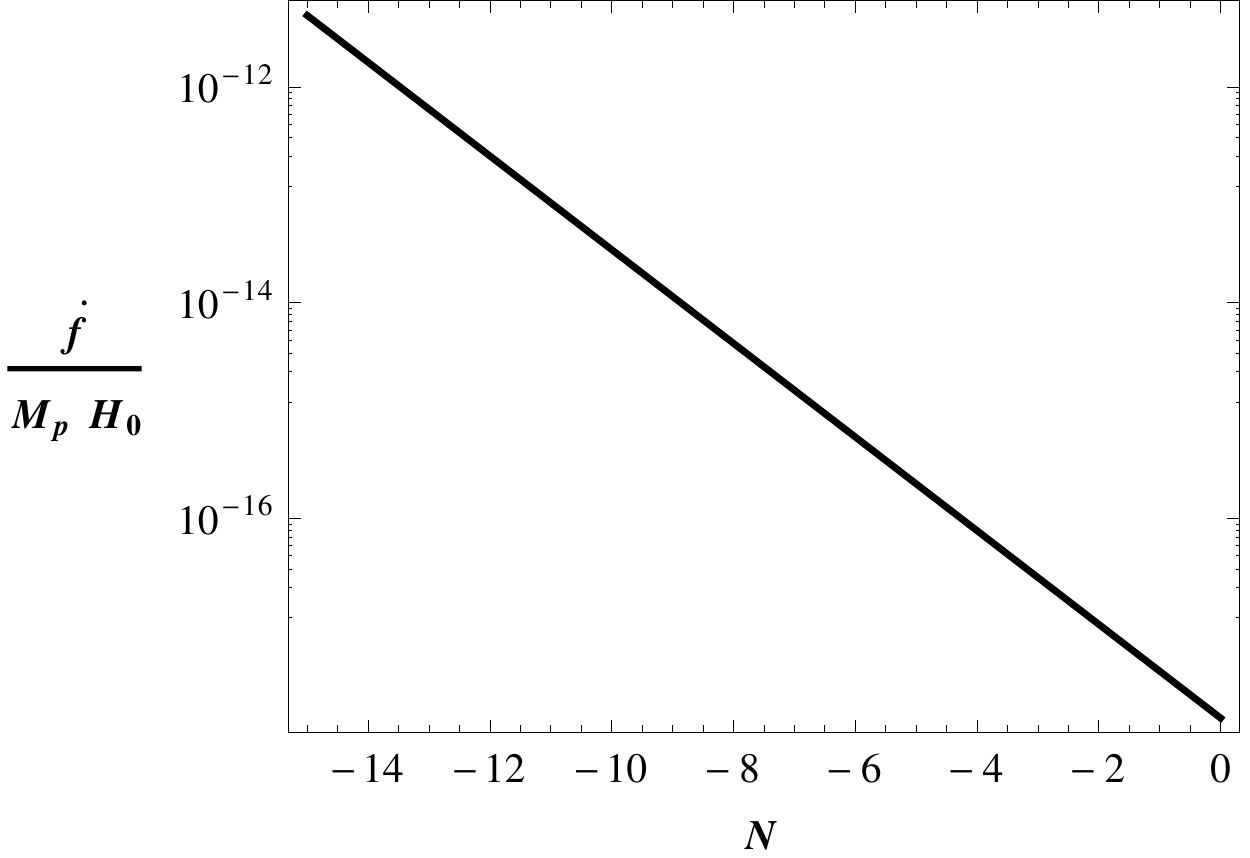}
  \caption{Velocity of gauge field $\frac{\dot f}{M_P H_0}$
   plotted against the number of e-foldings $N$. }
  \label{gauge-velo}
\end{figure}

\begin{figure}
\centering
\includegraphics[scale=0.5]{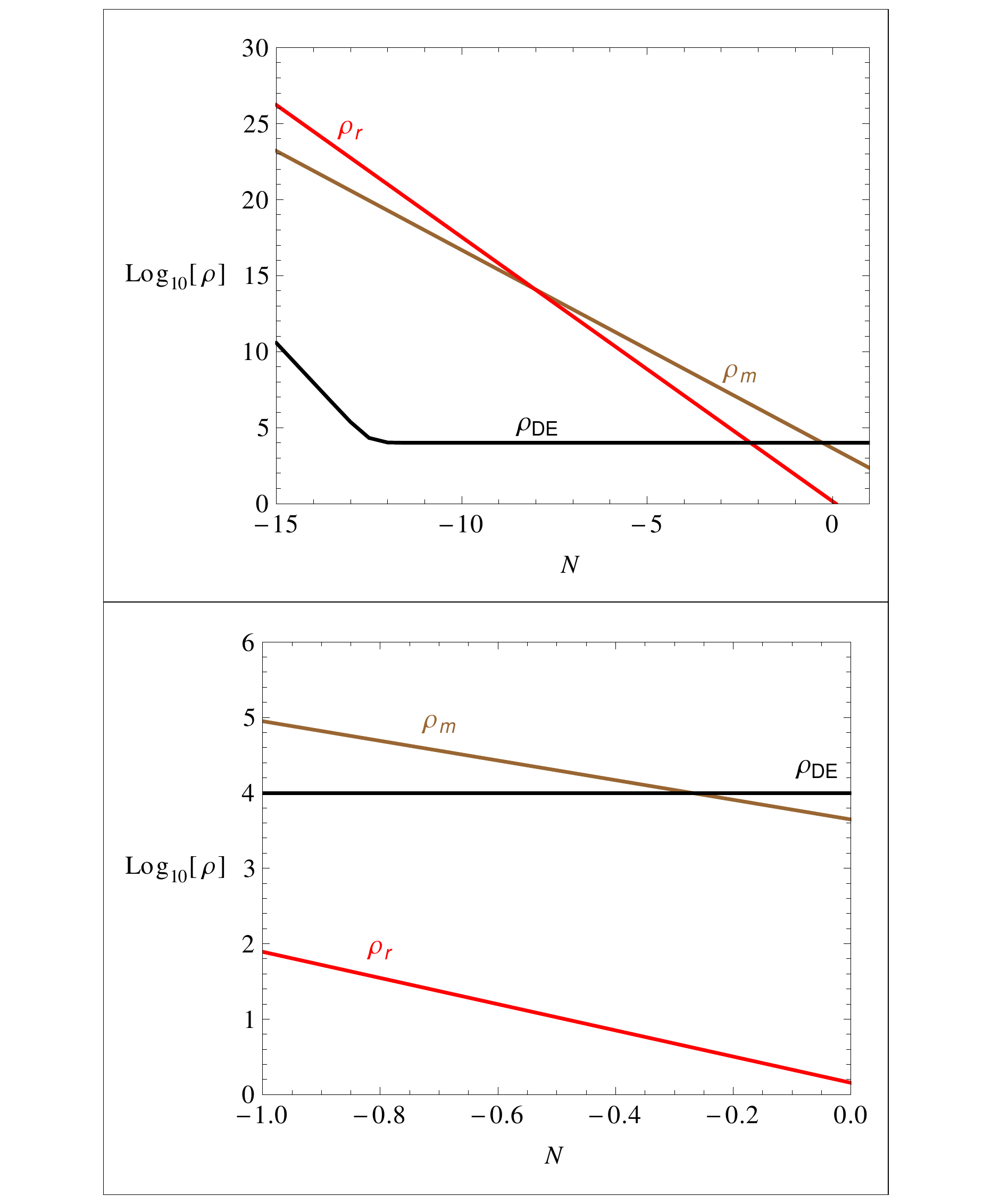}
\caption{Plot of the density in the Universe with number of e-foldings $N$ given by
$N=-\ln(1+z)$ where $z$ is the corresponding redshift.}
\label{rho}
\end{figure}

\begin{figure}
\centering
\includegraphics[scale=0.5]{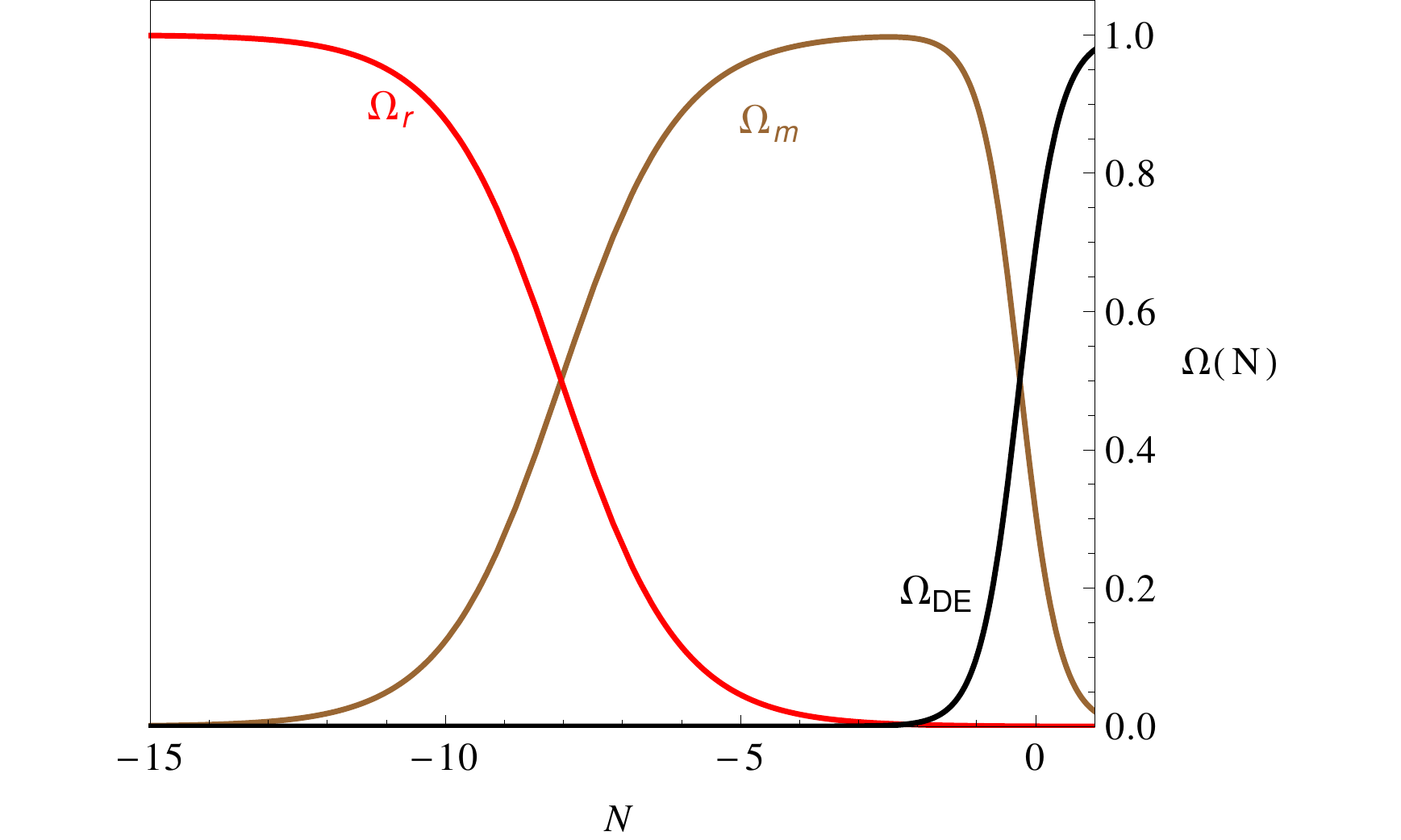}
\caption{Plot of the density parameters in the Universe with the  number of 
e-foldings $N$ given by $N=-\ln(1+z)$ where $z$ is the corresponding redshift.}
\label{omega}
\end{figure}

\begin{figure}
\centering
\includegraphics[scale=0.5]{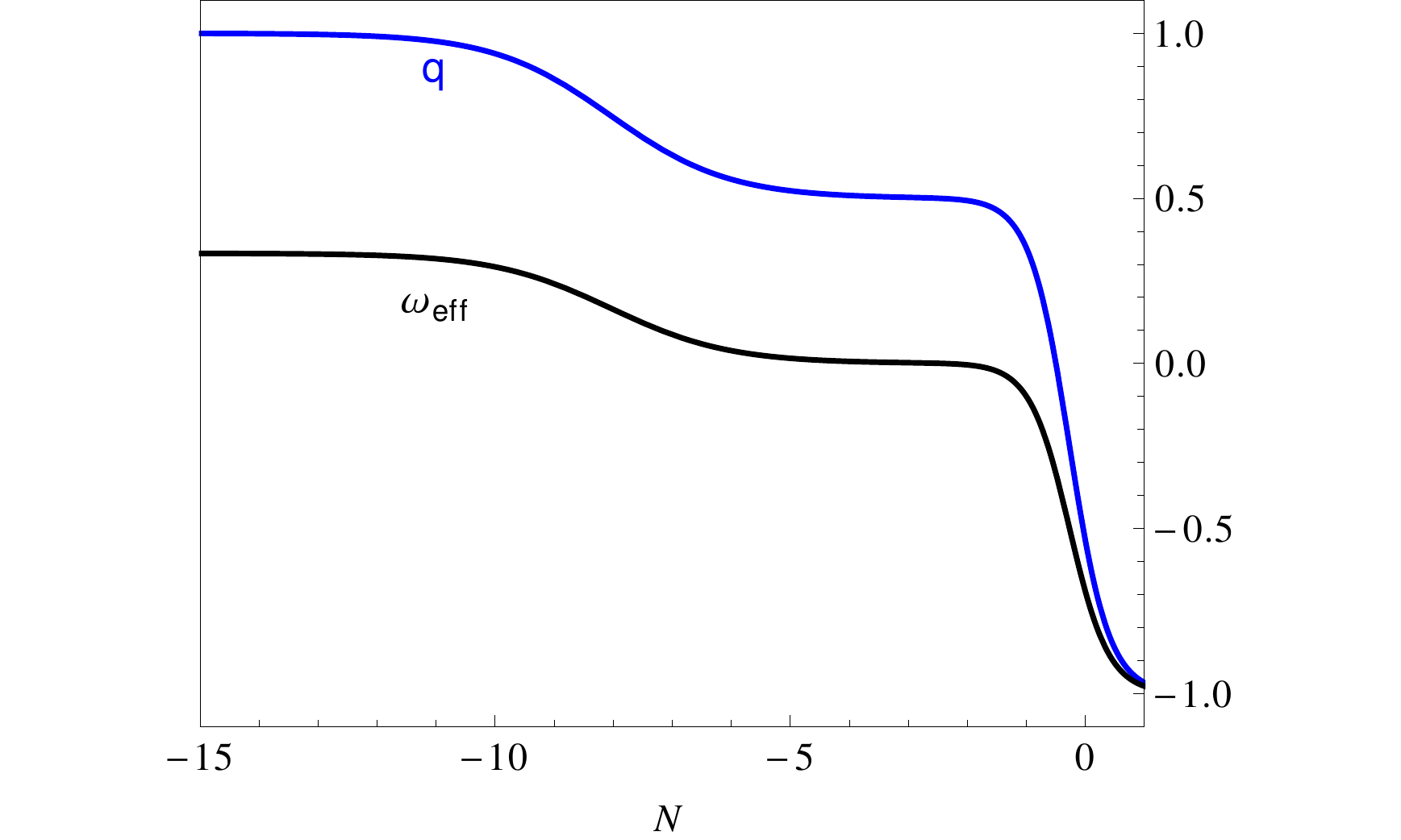}
\caption{Plot of deceleration parameter $q$ and effective equation of state of 
the universe $\omega_{\rm eff}$ with the number of e-foldings $N$ given by
$N=-\ln(1+z)$ where $z$ is the corresponding redshift.}
\label{qw}
\end{figure}

The dynamical analysis of the model reveals that there exists an attractor
solution for the model in $x_1=0,~y_1=0,~z_1=0,~x_2=0,~y_2=1,~r=0, ~m=0,
w_1=0$ plane for which the deceleration parameter turns out to be $q=-1$. This attractor sloution corresponds to late-time cosmic acceleration. 
In Fig. \ref{dyna}, the
dynamical evolution of the model is shown in the phase planes $(x_1,~y_1)$ and 
$(x_1,~y_2)$. 
In the Fig. \ref{dyna}(left) the system starts from
(1, 0) (solid red), (1, 1) (red dotdashed), (1, 2) (red dashed),
 (0.3, 2) (solid blue), (0, 2) (blue dashed), (-0.3, 2) (blue dotdashed)
(-1,2) (green dotdashed), (-1,1) (green dashed) and (-1,0)
(solid green)
in the 
$(x_1,~y_1)$ plane and approaches the attractor solution
(0, 0) as $N$ changes 
from $N=-15$ to $N=0$. Similarly, in the Fig. \ref{dyna}(right)
the system starts from 
(1, 1) (solid red), (1, 2) (red dashed), (1, 5)
 (red dotdashed), (0.5, 5) (solid blue), (0, 5) (blue dashed),
(-0.5,5) (blue dotdashed), (-1, 5) (green dashed), (-1, 2) 
(green dotdashed) and (-1,1) (solid green)  
in the 
$(x_1,~y_2)$ plane and approaches the attractor solution
(0, 1) as $N$ changes 
from $N=-15$ to $N=0$.
We solve the autonomous system for the initial conditions given by
$x_1=10^{-18}$, $y_1=10^{-18}$,
$z_1=10^{-18}$, $x_2=10^{-18}$, $y_2=0.831$,
$w_1=10^{2}$ and $r= 10^{-2}$ at $z=0$. 
These initial conditions chosen here, are in 
consideration with the Higgs vev ($v\sim 246$ GeV),
radiation density parameter at the present epoch of order $10^{-4}$
and the total length of the radiation dominated 
epoch (see \cite{Rinaldi:2014yta,Alvarez:2019ues} for
details).
The velocities of the Higgs field and the gauge field are shown in the 
 Figs. \ref{phi-velo} and \ref{gauge-velo}.
From the Figs. \ref{phi-velo} and \ref{gauge-velo}, 
it is evident that 
the Higgs and the gauge field are moving very slowly.
The reason behind this is the
presence of gauge field which introduces enough 
Hubble friction as well as flattens the mexican hat potential 
of Higgs field (Eq. \ref{higgs-eff-pot})
so as to make the Higgs field move very slowly along the 
potential. Later  
this fact will also be reflected in the Fig. \ref{rho} (top) where 
$\rho_{\rm DE}$ starts mimicking cosmological constant deep in the 
radiation era.
In the Fig. \ref{rho} (top), we show the variation of density of radiation, matter
and dark energy with the number of e-foldings $N$ and variation of 
the density parameters are shown in Fig. \ref{omega} for $\Omega_m^{(0)}=0.31$
and $H_0=69\,{\rm Km sec^{-1} Mpc^{-1}}$ and $\alpha=1$. 
From these two figures this is evident
that the dark energy dominates very recently (around $N\sim -0.25$ as
clearly shown in Fig. \ref{rho} (bottom) also). 
Moreover it is evident from Fig. \ref{rho} (top)
that the Higgs dark energy though varies 
initially but starts 
mimicking the cosmological constant around $N=-12$ i.e., well in the  radiation
dominated era.
The plot of the deceleration parameter
$q$ and the effective equation of state $\omega_{\rm eff}$
of the universe for all the components
i.e., radiation, matter and the dark energy are shown in Fig. \ref{qw}. The 
acceleration of the universe corresponds to $q<0$ and $\omega_{\rm eff}<-1/3$.

\section{Observational Constraints}
\label{Obs-cons}
In this era of precision cosmology models of dark energy are highly constrained. 
Passing the test of observational data only makes the model acceptable despite
their theoretical viability. In this section we describe the observational data that
are used to constrain the model parameters in this Einstein Yang-Mills
Higgs dark energy and the formalism for  the data analysis as well.

Supernovae Type Ia are accepted as the standard candles in astrophysical
observations. Incidentally it happened to be the first probe for the
discovery of the late time cosmic acceleration \cite{Perlmutter:1998np,Riess:1998cb}.
We consider here  279 Supernovae Type Ia (SNe Ia) observational data from  Pan-STARRS1 
Medium Deep Survey in the redshift range $0.03 < z < 0.68$ along with
the other SNe Ia data from Sloan Digital Sky Survey (SDSS)
\cite{Frieman_2007,Kessler_2009,Sako:2014qmj}, SNLS 
\cite{Conley_2010,2011ApJ...737..102S}, 
and ESSENCE  \cite{Miknaitis_2007,Wood_Vasey_2007,Narayan_2016} 
and SCP \cite{Suzuki:2011hu}. The combined
data set known as Pantheon Sample \cite{2018ApJ...859..101S}
consists  of 1048 SNe Ia data points in the redshift range $0.01 < z < 2.3$.
The distance modulus for type Ia supernova as a function of the redshift is given by
\bea
\mu(z) &=& 5 \log_{10} (D_L(z)) + \mu_0\,,
\eea
where $D_L(z)= H_0 d_L(z)/c$ ($c$ is speed of light in free space) and
$\mu_0=42.38-5\log_{10} h$ for $H_0=100h\,{\rm Km\,Sec^{-1}\,Mpc^{-1}}$. 
The chi-square for supernovae data is defined as
\bea
\chi^2_{\rm SN}(p_s) &=& \sum_i \left[\frac{\mu_{\rm obs}(z_i)-
\mu_{\rm th}(z_i,p_s,\mu_0)}
{\sigma_i}\right]^2\,,
\eea
where $p_s$ are the model parameters and $z_i$ are the redshifts of the observational
supernovae type Ia data. $\mu_{\rm obs}$ and $\mu_{\rm th}$
are the observational and theoretical distance modulus respectively.
The chi-square is marginalised over the nuisance parameter
$\mu_0$ \cite{Lazkoz:2005sp} 
and the marginalised chi-square is given by 
\bea
\chi^2_{\rm SN} &=& A(p_s) - \frac{B(p_s)^2}{C}\,,
\eea
where $A$, $B$ and $C$ are given by
\bea
A(p_s)&=& \sum_i \left[\frac{\mu_{\rm obs}(z_i)-
\mu_{\rm th}(z_i,p_s,\mu_0)}
{\sigma_i}\right]^2\,,\\
B(p_s)&=& \sum_i \frac{\mu_{\rm obs}(z_i)-
\mu_{\rm th}(z_i,p_s,\mu_0)}
{\sigma_i^2}\,,\\
C&=& \sum_i \frac{1}{\sigma_i^2}\,,
\eea
Cosmic microwave background shift parameter $R$  is a model independent 
parameter that can also be used to constrain the models of dark energy. It is obtained
from the first peak of temperature anisotropy plot of the cosmic microwave
background radiation. The CMBR shift parameter is defined as
\bea
R(z_*) &=& (\Omega_m^0 H_0^2)^{1/2} \int _0^{z_*} \frac{dz}{H(z)} \,,
\eea
where $z_*$ corresponds to the redshift of the radiation matter decoupling epoch.
The chi-square for CMBR shift parameter is defined as
\bea
\chi^2_{\rm CMB} &=& \left[\frac{R_{\rm th }(z_*,p_s)-R_{\rm obs}(z_*)}{\sigma_R}
\right]^2\,.
\eea
Needless to mention that $p_s$ are the model parameters. We use the CMBR
shift parameter from latest  Planck observations $R=1.7499\pm 0.0088$ at the
redshift of decoupling era $z_*=1091.41$ \cite{Ade:2013zuv}.

Observational Hubble data is a direct measurement of expansion rate of universe
with the redshifts. It is another tool to constrain the dark energy models.
The chi-square for observational Hubble data is given by
\bea
\chi^2_{\rm OHD} &=& \sum_i 
\left[\frac{H_{\rm obs}(z_i)- H_{\rm th}(z_i,p_s)}{\sigma_i^2}\right]^2\,,
\eea
We use the 31 data points of $H(z)$ 
for the purpose of $\chi^2_{\rm OHD}$
analysis. 
\begin{table}
	\centering
\begin{tabular}{|c|c|c|}
\hline
$z$ & $H(z)$ & $\sigma_{H(z)}$ \\
	&${\rm  KmSec^{-1} Mpc^{-1}}$ & ${\rm KmSec^{-1} Mpc^{-1}}$\\
\hline
 0.07 & 69.0 & 19.6  \cite{2014RAA....14.1221Z}  \\
 0.09 & 69.0 & 12.0  \cite{Simon:2004tf}  \\
 0.12 & 68.6 & 26.2  \cite{2014RAA....14.1221Z}  \\
 0.17 & 83.0 & 8.0  \cite{Simon:2004tf}    \\
 0.179 & 75.0 & 4.0  \cite{2012JCAP...08..006M}\\
 0.199 & 75.0 & 5.0  \cite{2012JCAP...08..006M}\\
 0.2 & 72.9 & 29.6  \cite{2014RAA....14.1221Z}  \\
 0.27 & 77.0 & 14.0  \cite{Simon:2004tf} \\
 0.28 & 88.8 & 36.6  \cite{2014RAA....14.1221Z} \\
 0.352 & 83.0 & 14.0  \cite{2012JCAP...08..006M} \\
 0.3802 & 83.0 & 13.5  \cite{Moresco:2016mzx} \\
 0.4 & 95.0 & 17.0  \cite{Simon:2004tf}  \\ 
 0.4004 & 77.0 & 10.2  \cite{Moresco:2016mzx}  \\
 0.4247 & 87.1 & 11.2   \cite{Moresco:2016mzx} \\
 0.4497 & 92.8 & 12.9   \cite{Moresco:2016mzx}  \\
 0.47 & 89.0 & 49.6   \cite{Ratsimbazafy:2017vga} \\
0.4783 & 80.9 & 9.0   \cite{Moresco:2016mzx}\\
0.48 & 97.0 & 62.0  \cite{2010JCAP...02..008S}  \\
0.593 & 104.0 & 13.0  \cite{2012JCAP...08..006M} \\
0.68 & 92.0 & 8.0  \cite{2012JCAP...08..006M} \\
0.781 & 105.0 & 12.0   \cite{2012JCAP...08..006M} \\
 0.875 & 125.0 & 17.0  \cite{2012JCAP...08..006M} \\
0.88 & 90.0 & 40.0  \cite{2010JCAP...02..008S} \\
0.9 & 117.0 & 23.0  \cite{Simon:2004tf} \\
1.037 & 154.0 & 20.0  \cite{2012JCAP...08..006M} \\
1.3 & 168.0 & 17.0  \cite{Simon:2004tf}\\
1.363 & 160.0 & 33.6  \cite{Moresco:2015cya}\\
1.43 & 177.0 & 18.0  \cite{Simon:2004tf} \\
1.53 & 140.0 & 14.0  \cite{Simon:2004tf} \\
1.75 & 202.0 & 40.0  \cite{Simon:2004tf} \\
1.965 & 186.5 & 50.4  \cite{Moresco:2015cya} \\
 \hline
\end{tabular}
	\caption{The 31 $H(z)$ data points \cite{Yang:2019fjt}.}
	\label{hub-data}
\end{table}
The measurements of observational Hubble data are summarized in Tab. \ref{hub-data}
\cite{Yang:2019fjt}.

Before the recombination epoch the baryons were tightly coupled to the photons
and as a result of this tight coupling the acoustic oscillations created small density 
fluctuations in baryon photon plasma. 
In the expanding universe, this density fluctuations left
an imprint in the large scale structures which provides a standard ruler in cosmology.
Baryon acoustic oscillation is the powerful tool for constraining dark energy models.
The sound horizon at a redshift $z_d$ for drag epoch is given by
\bea
r_d &=& \frac{c}{\sqrt{3}}\int_{z_d}^{\infty} 
\frac{dz}{\sqrt{1+\frac{3\Omega_b^{(0)}}{4\Omega_\gamma^{(0)}}\frac{1}{1+z}}H(z)}\,,
\eea
where the drag redshift $z_d$ is given by
\bea
z_d &=& \frac{1291 \left(\Omega_m^{(0)} h^2\right)^{0.251}}
{1+0.659\left(\Omega_m^{(0)} h^2\right)^{0.828}}
\left[1+b_1\left(\Omega_b^{(0)} h^2\right)^{b_2}\right]\,,
\eea
with 
\bea
b_1&=& 0.313\left(\Omega_b^{(0)} h^2\right)^{-0.419}
\left[1+0.607\left(\Omega_m^{(0)} h^2\right)^{0.674}\right]\,,\\
b_2 &=& 0.238 \left(\Omega_m^{(0)} h^2\right)^{0.223}\,,
\eea
and $\Omega_b^{(0)} h^2 = 0.02236$, $\Omega_\gamma^{(0)} h^2=2.469\times 10^{-5}$
\cite{Aghanim:2018eyx}.
In a spatially flat universe the angular diameter distance $D_A(z)$, 
the Hubble distance $D_H(z)$ and the effective distance $D_V(z)$
are respectively given by,
\bea
D_A(z) &=& \frac{c}{(1+z)}\int_0^z \frac{dz}{H(z)}\,,\\
D_H(z) &=& \frac{c}{H(z)}\,,\\
D_V(z) &=& \left[\left(\frac{d_L(z)}{1+z}\right)^2 \frac{c z}{H(z)}\right]^{1/3}
\eea
where $c$ the speed of light in vacuum.
Here we use both the isotropic and anisotropic BAO data that are tabulated in Tabs.
\ref{bao-iso} and \ref{bao-aniso} \cite{Evslin:2017qdn}. 
\begin{table}
  \centering
  \begin{tabular}{|ccc|}
  \hline
    Data set & Redshift & $D_V(z)/r_d$ \\
    \hline
    6dF & z=0.106 & $2.98\pm0.13$  \cite{Beutler:2011hx} \\
    MGS & z=0.15 & $4.47\pm0.17$  \cite{Ross:2014qpa} \\
    eBOSS quasars & z=1.52 & $26.1\pm1.1$  \cite{Ata:2017dya} \\
    \hline
  \end{tabular}
  \caption{Isotropic BAO data.}
\label{bao-iso}
\end{table}
\begin{table}
  \centering
	\begin{adjustbox}{width=1.0\columnwidth}
	\begin{tabular}{|ccc|}
    \hline
    Data set & Redshift & $D_{A/H}(z)/r_d$ \\
    \hline
    BOSS DR12 & z=0.38 & $7.42 (A)$  \cite{Alam:2016hwk} \\
    BOSS DR12 & z=0.38 & $24.97 (H)$  \cite{Alam:2016hwk} \\
    BOSS DR12 & z=0.51 & $8.85 (A)$  \cite{Alam:2016hwk} \\
    BOSS DR12 & z=0.51 & $22.31 (H)$  \cite{Alam:2016hwk} \\
    BOSS DR12 & z=0.61 & $9.69 (A)$  \cite{Alam:2016hwk} \\
    BOSS DR12 & z=0.61 & $20.49 (H)$  \cite{Alam:2016hwk} \\
    BOSS DR12 & z=2.4 & $10.76 (A)$  \cite{Bourboux:2017cbm} \\
    BOSS DR12 & z=2.4 & $8.94 (H)$  \cite{Bourboux:2017cbm} \\
    \hline
	\end{tabular}
	\end{adjustbox}
  \caption{Anisotropic BAO data. }
\label{bao-aniso}
\end{table}
The covariance matrix ${\bm {C}}$ 
associated with
the anisotropic BAO measurements is given by
\begin{adjustbox}{width=0.62\columnwidth}
	\begin{equation*}
	\bm{C}=
    \begin{pmatrix}
      0.0150 & -0.0357 & 0.0071 & -0.0100 & 0.0032 & -0.0036 & 0 & 0 \\
      -0.0357 & 0.5304 & -0.0160 & 0.1766 & -0.0083 & 0.0616 & 0 & 0 \\
      0.0071 & -0.0160 & 0.0182 & -0.0323 & 0.0097 & -0.0131 & 0 & 0 \\
      -0.0100 & 0.1766 & -0.0323 & 0.3267 & -0.0167 & 0.1450 & 0 & 0 \\
      0.0032 & -0.0083 & 0.0097 & -0.0167 & 0.0243 & -0.0352 & 0 & 0 \\
      -0.0036 & 0.0616 & -0.0131 & 0.1450 & -0.0352 & 0.2684 & 0 & 0 \\
      0 & 0 & 0 & 0 & 0 & 0 & 0.1358 & -0.0296 \\
      0 & 0 & 0 & 0 & 0 & 0 & -0.0296 & 0.0492 \\
    \end{pmatrix}\,.
	\end{equation*}
    \end{adjustbox}

The total chi-square for isotropic and anisotropic BAO data is given by
\bea
\chi^2_{\rm BAO} &=& \chi_{\rm iso}^2+ \chi_{\rm aniso}^2\,,
\eea
where
\bea
\chi_{\rm iso}^2 &=& \sum_i 
\left[\frac{D_V(z_i)/r_d-D_V(z_i,p_s)/r_d}{\sigma_i}\right]^2\,,\\
\chi_{\rm aniso}^2 &=& X_{\rm aniso}^T {\bm C}^{-1} X_{\rm aniso} 
\eea
where $X_{\rm aniso}$ is column matrix given by,    
\begin{equation}
X_{\rm aniso}=\left( \begin{array}{c}
        \frac{D_A(0.38)}{r_d} - 7.42 \\
        \frac{D_H(0.38)}{r_d} - 24.97 \\
        \frac{D_A(0.51)}{r_d} - 8.85 \\
        \frac{D_H(0.51)}{r_d} - 22.31 \\
        \frac{D_A(0.61)}{r_d} - 9.69 \\
        \frac{D_H(0.61)}{r_d} - 20.49 \\
        \frac{D_A(2.4)}{r_d} - 10.76 \\
        \frac{D_H(2.4)}{r_d} - 8.94 
        \end{array} \right)\,.
\end{equation}

The total combined chi-square for all the aforesaid data sets 
i.e., SNe Ia, CMB shift parameter, OHD, BAO is given by,
\bea
\chi^2_{\rm tot} &=& \chi^2_{\rm SN} + \chi^2_{\rm OHD} + \chi^2_{\rm BAO} +
\chi^2_{\rm CMB} \,.
\label{chitot}
\eea
We use this total chisquare defined in Eq. (\ref{chitot}) for the
data analysis purpose of the
Yang-Mills Higgs dark energy model and constrain the parameters space.

\begin{figure}
\centering
\includegraphics[scale=0.6]{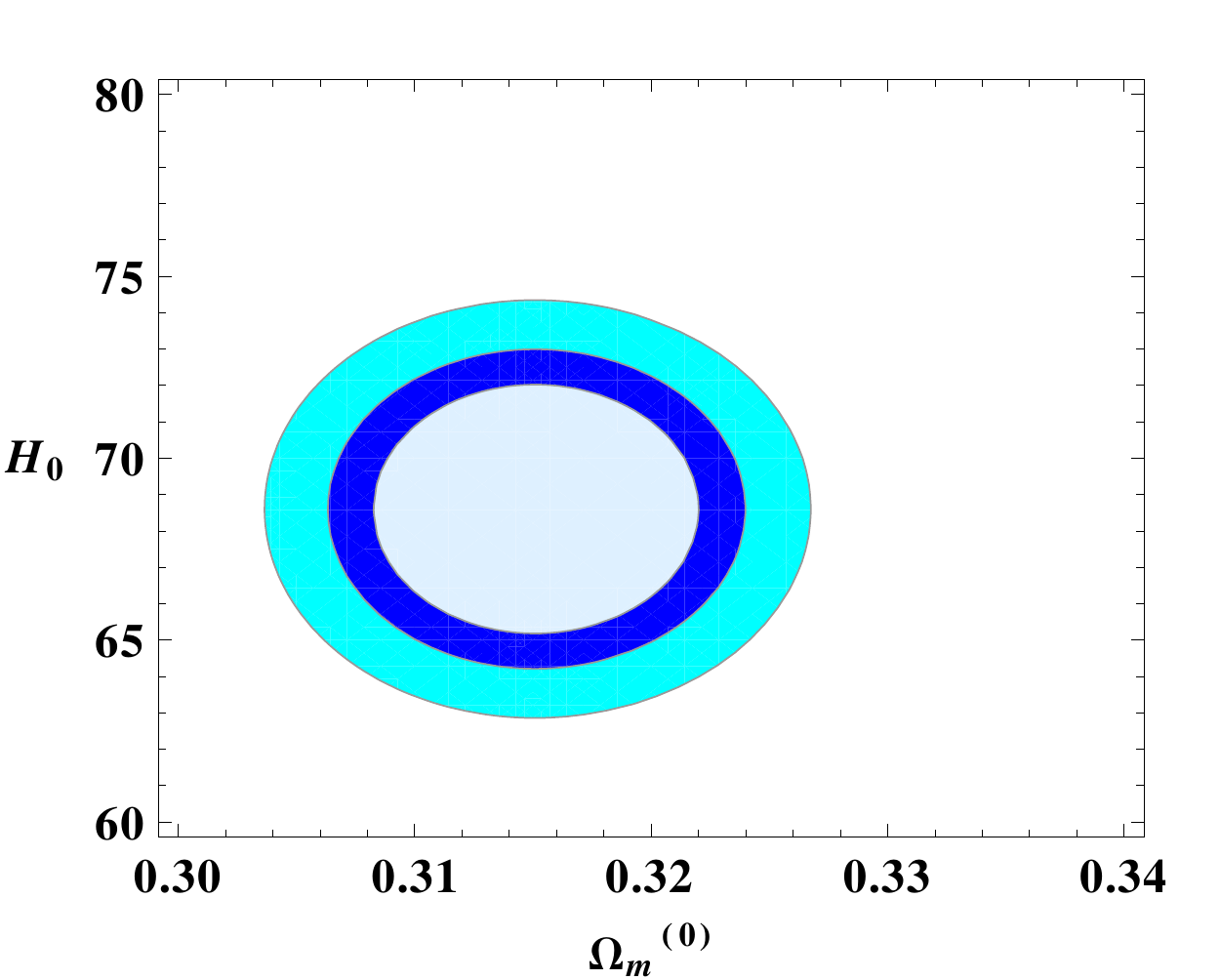}
\caption{Observational constraints on the parameters space ($\Omega_m^{(0)}-H_0$)
at the 68.3\% (light blue), 90\% (dark blue) and 99\% (cyan) confidence levels.}
\label{omegam-H0}
\end{figure}

\begin{figure}
\centering
\includegraphics[scale=0.6]{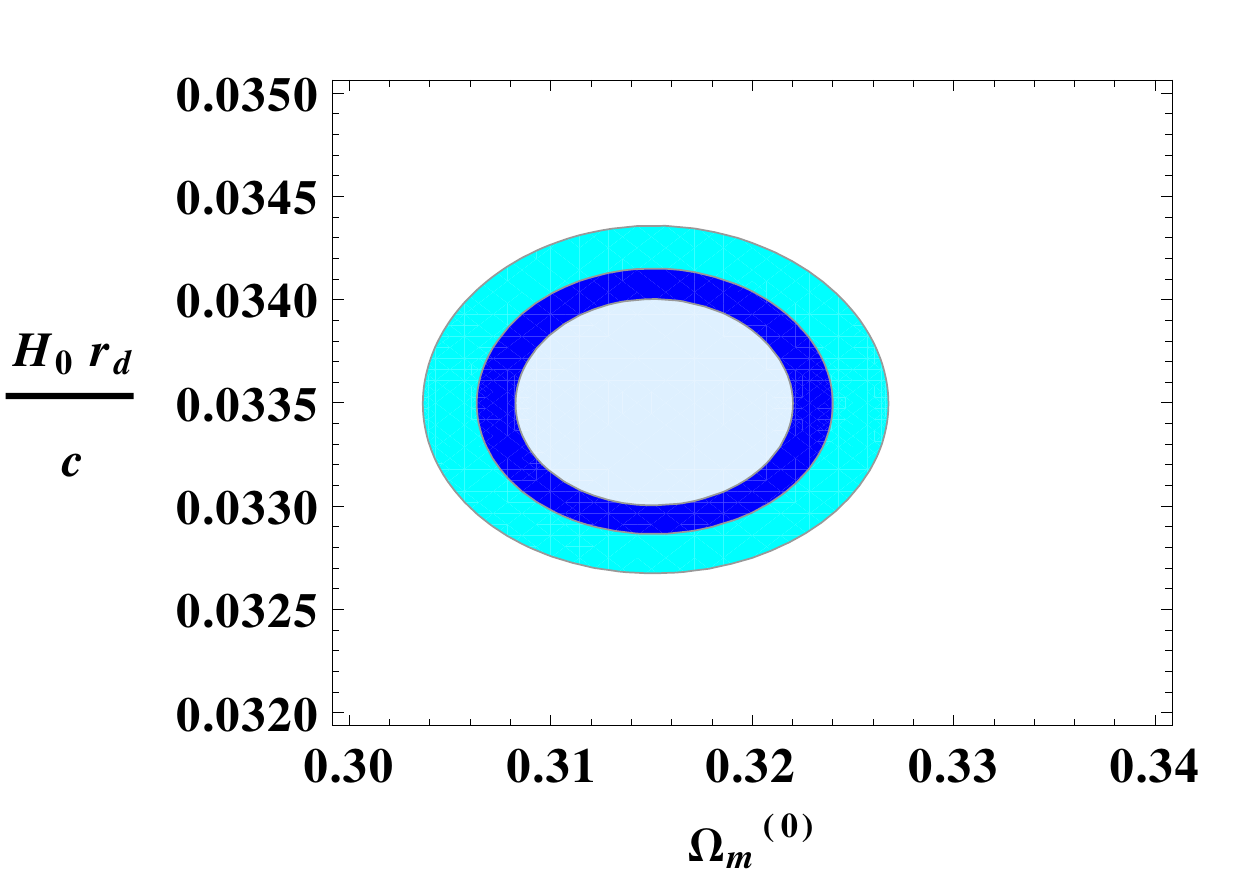}
\caption{Observational constraints on the parameters space ($\Omega_m^{(0)}-H_0r_d/c$)
at the 68.3\% (light blue), 90\% (dark blue) and 99\% (cyan) confidence levels.}
\label{omegam-rd}
\end{figure}

\begin{figure}
\centering
\includegraphics[scale=0.6]{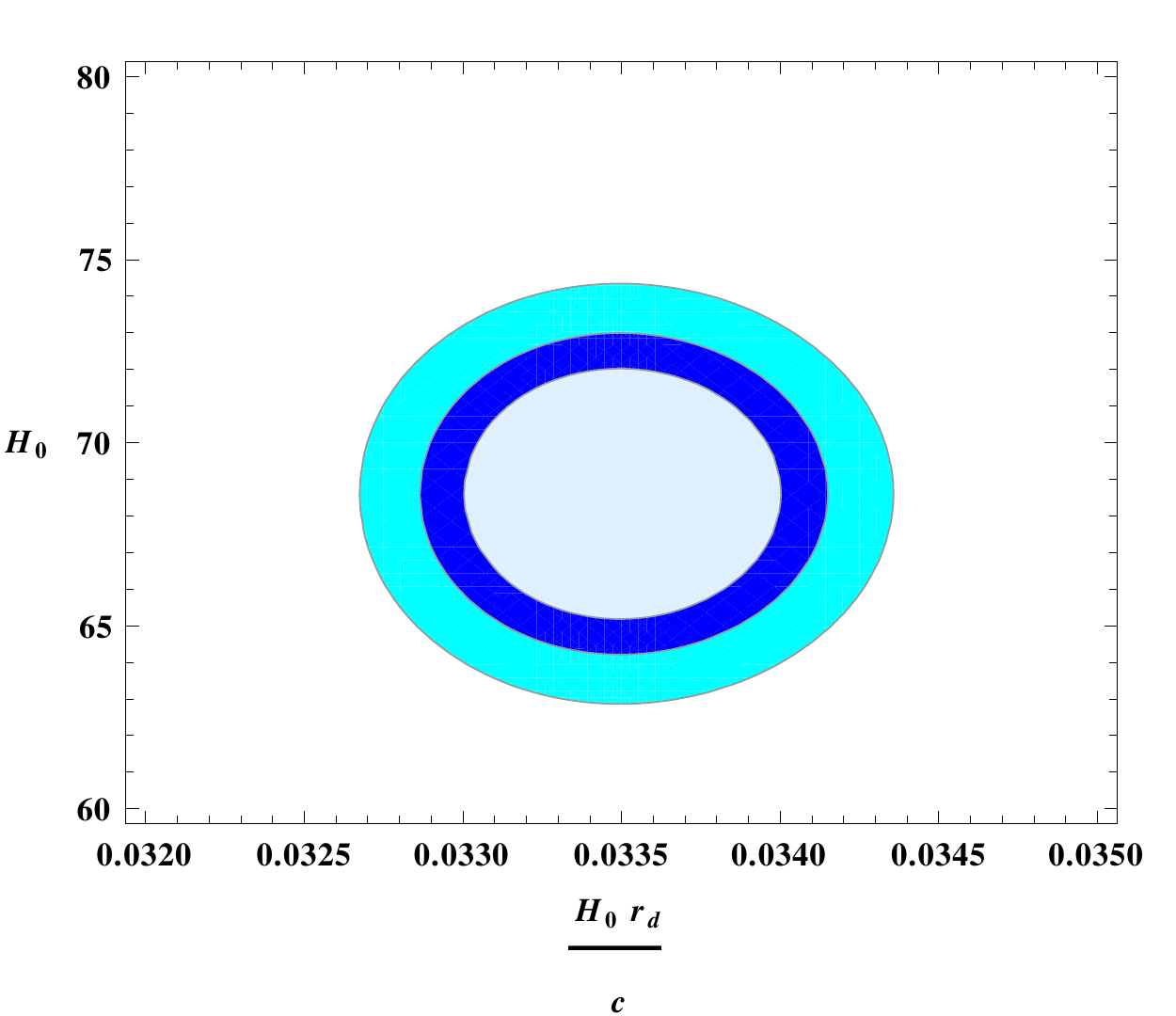}
\caption{Observational constraints on the parameters space ($H_0r_d/c-H_0$)
at the 68.3\% (light blue), 90\% (dark blue) and 99\% (cyan) confidence levels.}
\label{H0-rd}
\end{figure}

\begin{figure}
\centering
\includegraphics[scale=0.6]{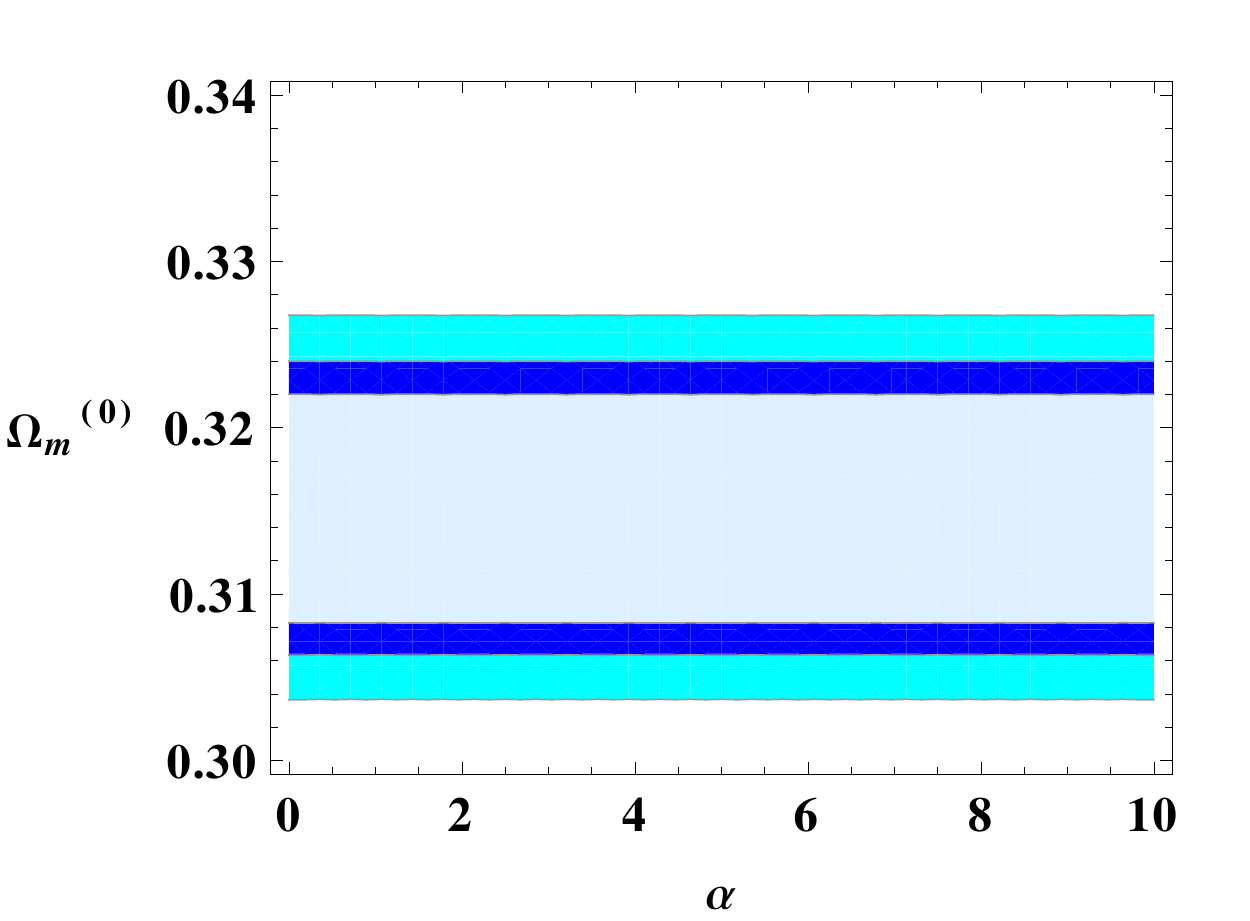}
	\caption{Observational constraints on the parameters space 
	($\alpha- \Omega_m^{(0)}$)
at the 68.3\% (light blue), 90\% (dark blue) and 99\% (cyan) confidence levels.}
\label{alpha-om}
\end{figure}

In what follows, we describe the model parameters and the results of the 
chi-square analysis of the observational data. 
In this Higgs dark energy model we consider four parameters namely
$\alpha$, $\Omega_m^{(0)}$, $H_0$ and $H_0 r_d/c$ to fit the chi-square
with the latest observational data from SNe Ia, OHD, BAO and CMB.
In the Fig. \ref{omegam-H0}, we present the 68.3\%, 90\% and 99\% confidence level plot 
for the parameters $\Omega_m^{(0)}$ and $H_0$ with the contour shadding by the 
light blue, dark blue and cyan colours respectively. 
The total chi-square turns out
to have a minima at $\Omega_m^{(0)} \simeq 0.315$ and 
$H_0 \simeq 68.6 \,{\rm KmSec^{-1} Mpc^{-1}}$ and $H_0 r_d/c\simeq 0.0335$ which 
with the best-fit  value of $H_0$ and speed of light in vacuum gives 
$r_d\simeq 146.5$ Mpc.
Fig. \ref{omegam-rd} and \ref{H0-rd} shows the observationally allowed parameters 
space in $\Omega_m^{(0)}-H_0r_d/c$ and $H_0r_d/c-H_0$ at 68.3\%, 90\% and 99\%
confidence levels with the same colours mentioned above.
It is worth mentioning here that all this confidence contours corresponds to value of 
$\alpha=1$ \cite{Orjuela-Quintana:2020klr}. 
The parameter $\alpha$ cannot be constrained from the present 
observational data we have considered here. A confidence contour is shown in 
Fig. \ref{alpha-om} in the $\alpha-\Omega_m^{(0)}$ parameters space from where
it is evident that the 
present data is unable to put any bound on the parameter $\alpha$.
With a total of 1091 data points from SNe Ia, OHD, BAO and CMB, we find 
from our data analysis a chi-square
per degrees of freedom to be around 0.983 i.e., very close to 1
which in turn refelects 
the fact that the fitment of the model parameters 
are in good agreement with the observational data sets \cite{Bevington:1305448}.

\section{Discussions and Conclusions}
\label{conc}

The present work shows that the dark energy can have its origin 
in the particle physics sector consistent with the observations 
coming from particle physics, such as Higgs vev and the dimensionless 
coupling constant arising from the SU(2) gauge coupling constant and 
the quartic coupling constant of Higgs field, as well as from the 
cosmological observations such as SNe Ia, OHD, BAO, CMB shift parameter data.
As the masses of the elementary particles are generated via the interactions with
the Higgs field, it is assumed that the Higgs field is isotropically and homogeneously
present in the universe. However the mexican hat potential of Higgs field is not flat
enough to lead to the recent cosmic acceleration. The presence of SU(2) gauge field enhances the
Hubble friction as well as generates an effective flat enough potential so that the
Higgs dark energy can give rise to late time cosmic acceleration. The model
turns out to be a viable dark energy model having its origin in the standard model of elementary particles
so far as the observations are concerned. This is worth mentioning here that 
spacetime symmetries of FLRW metric leaves us with the only degree of freedom 
$f(t)$ allowed from the SU(2) gauge sector. From the analysis it is also evident that an 
effective potential for gauge field arises due to the interaction with the Higgs field. However
the effective potential for the gauge degrees of freedom is minimum at $f=0$ thereby leading to the 
vanishing vacuum expectation value to the gauge fields in this theory \cite{Alvarez:2019ues}
which is completely in agreement with the usual assumptions for the gauge fields in standard model 
of particle physics.

In this work, we study Higgs dark energy model in presence of gauge field 
in light of observational data from supernovae type Ia, baryon acoustic oscillation,
observational Hubble data and cosmic microwave background shift parameter data.
In performing the data analysis, we considered the initial conditions 
at the present epoch for
dynamical evolution of the autonomous system. The choice of initial condition
is in consideration with the vacuum expectation value of Higgs
$v \sim$ 246 GeV, radiation density parameter at the present epoch 
of the order $10^{-4}$
and the total length of the radiation dominated epoch
that leads to initial values of $x_1=10^{-18}$, $y_1=10^{-18}$, 
$z_1=10^{-18}$, $x_2=10^{-18}$, $y_2=0.831$,
$w_1=10^{2}$ and $r= 10^{-2}$ at $z=0$ i.e., present epoch 
\cite{Rinaldi:2014yta,Alvarez:2019ues}. These choice of initial conditions
lead to correct cosmological dynamics for the observational universe
as evident from Fig \ref{rho} and the cosmic acceleration is a recent
phenomenon. Moreover from the same figure it appears that the Higgs dark energy
starts mimicking cosmological constant well in the radiation dominated era.
The chi-square analysis of the observational data significantly constrains  the
model parameters ($\alpha,~\Omega_m^{(0)},~H_0, ~H_0 r_d/c$). The minimum
combined chi-square for all the data sets is obtained at the parameter
values ($\Omega_m^{(0)},~H_0, ~H_0 r_d/c$) $\sim$ (0.315, 68.6, 0.0335). Hence 
the sound horizon at the redshift of drag epoch turns out to be around 146.5 Mpc
which is in remarkably good agreement with the Planck 2018 results 
\cite{Aghanim:2018eyx}. Also this is worth mentioning here that the chi-square
per degrees of freedom is slightly greater than 0.98 which is the indication
of a good fitting of the model with the observational data 
\cite{Bevington:1305448}. However data
is still unable to provide any constraint on the model parameter $\alpha$
as is evident from Fig. \ref{alpha-om}. Needless to mention that 
the Higgs or the gauge field in the theory being minimally coupled to gravity
does not conflict with the observational evidences of gravitational wave detection 
\cite{PhysRevLett.119.161101,Goldstein_2017,Abbott_2017b}. Thus the Higgs  field
in presence of gauge field turns out to be a viable candidate for 
dark energy so far as the observational data are concerned.

\section{Data Availability Statement}
All the data used in this article are available in the article except for
the Pantheon Sample SNe Ia data points 
\cite{2018ApJ...859..101S} which are available at 
http://archive.stsci.edu/\\doi/resolve/resolve.html?doi=10.17909/T95Q4X.
\bibliographystyle{spphys}
\bibliography{refs.bib}
\end{document}